\documentclass[11pt,a4paper]{article}
\usepackage{a4wide}
\usepackage{latexsym}
\usepackage{amssymb}
\usepackage{graphicx}
\usepackage{amsmath,cite}

 \makeatletter
\@addtoreset{equation}{section} \makeatother

\def\bfone{\relax{\rm 1\kern-.35em 1}}

 \makeatletter
\@addtoreset{equation}{section} \makeatother

\def\be {\begin{equation}}
\def\ee {\end{equation}}
\def\bea {\begin{eqnarray}}
\def\eea {\end{eqnarray}}
\def\bc {\begin{center}}
\def\ec {\end{center}}
\def\a  {\alpha}

\def\bfg {\begin{figure}}
\def\efg {\end{figure}}
\def\bi {\begin{itemize}}
\def\ei {\end{itemize}}
\def\nn {\nonumber}

\def\la {\label}
\def\le {\left}
\def\ri {\right}

\DeclareFontFamily{U}{rsf}{} \DeclareFontShape{U}{rsf}{m}{n}{
  <5> <6> rsfs5 <7> <8> <9> rsfs7 <10-> rsfs10}{}
\DeclareMathAlphabet\Scr{U}{rsf}{m}{n}

\begin{document}

\begin{center}
{\bf \large{ Towards a Cosmology with Minimal Length and Maximal Energy \\[10mm]}}
\large Ahmed Farag Ali$^{
\dagger \ddagger}$\footnote{\upshape{Electronic address~:~ahmed.ali@fsc.bu.edu.eg~;~afarag@zewailcity.edu.eg}} and  Barun Majumder $^{\star}$\footnote{Electronic address~:~\upshape{barunbasanta@iitgn.ac.in }}
\\[7mm]
\small{$^\ddagger$Centre for Fundamental Physics, Zewail City of Science and Technology}\\ \small{ Sheikh Zayed, 12588, Giza, Egypt.}\\[4mm]
\small{$^{\dagger}$Department of Physics, Faculty of Science, Benha University, Benha 13518, Egypt.}\\[4mm]
\small{$^\star$Indian Institute of Technology Gandhinagar, Ahmedabad, India.}
\end{center}

\begin{abstract}
The Friedmann-Robertson-Walker (FRW) universe and Bianchi I,II universes are investigated in the framework of the
generalized uncertainty principle (GUP) with a linear and a quadratic term in Planck length and momentum, which predicts minimum measurable length as well as maximum measurable momentum.
We get a dynamic cosmological bounce for the FRW universe. With Bianchi universe, we found that the universe may be still isotropic  by implementing GUP. Moreover, the wall velocity appears to be stationary with respect to the universe velocity which means that when the momentum of the Universe evolves into a maximum measurable energy, the bounce is enhanced against the wall which means no maximum limit angle is manifested anymore.

\end{abstract}




\section{Introduction}

The existence of space-like singularities is a generic feature of Einstein's
general theory of relativity and this is proved by the singularity theorems of Penrose,
Hawking and Geroch\cite{hawell}. The singularities are usually characterized
by divergences of curvature invariants or breakdown of geodesics. This indicates
a limit beyond which general theory relativity is not applicable anymore. It is widely
believed that a theory of quantum gravity will be the one which can provide insights on
the resolution of singularities. In fact there are
different approaches for quantum theory of gravity,
so the description at low energy limit also have competing candidates
even at the phenomenological level.\par
Various approaches for quantum gravity such as string theory and black hole physics,
have predicted the existence of a
minimum measurable length and an essential
modification of the Heisenberg uncertainty principle
\cite{guppapers, BHGUP, kmm, kempf, brau, Scardigli} to the so-called generalized
uncertainty principle (GUP). The GUP is based on the modification
of the fundamental commutation relation mainly in position and momenta.
Recent developments suggests that the minimal length scale can be implemented in
quantum mechanics and quantum field theory through the GUP.
Within the framework of the GUP  minimal length scale has been studied
in details in quantum mechanics, quantum electrodynamics, thermodynamics, black-hole
physics and cosmology. For some recent reviews on the phenomenology of different
minimal length scale scenarios inspired by various approaches to the quantum
gravity see \cite{HossenNico}. \par
Quite recently Ali et al. \cite{advplb, Das:2010zf} introduced a new approach which
predicts maximum observable momenta besides a minimal measurable length. This model
is built to be consistent with DSR\cite{cg}, string theory and black hole physics
\cite{guppapers, BHGUP}. Furthermore, it ensures that $[x_i,x_j]=0=[p_i,p_j]$
through Jacobi identity\cite{Ali:2011fa}. Accordingly, the modification of the Heisenberg
uncertainty relation near the Planck scale reads

\bea
[x_i,p_j] &=& i \hbar \left[\delta_{i j} - \alpha \left(p
\delta_{i j} + \frac{p_i p_j}{p}\right) +  \alpha^2 \left(p^2
\delta_{i j} + 3 p_i p_j\right) \right]~. \label{comm01}
\eea

Here $\alpha$ is the deformation parameter given by
$\alpha=\alpha_0/(c\, M_{pl}) =\alpha_0 l_{pl}/\hbar$. $c$, $M_{pl}$ and $l_{pl}$
are speed of light in vacuum, Planck mass and Planck length respectively.
$\alpha_0$ is a dimensionless parameter usually considered to be of order unity.
The upper bounds on the parameter $\alpha_0$ have been estimated in \cite{Ali:2011fa}
and it was suggested that it could predict an intermediate length scale between
Planck scale and electroweak scale. A recent proposal suggested that these bounds
can be measured using quantum optics techniques in \cite{Nature,NatureGRW} which may be
considered as a milestone in quantum gravity phenomenology. Due to the GUP as
proposed in \cite{advplb} the physical momentum is redefined. As a result the
classical Hamiltonian as well as the quantum Hamiltonian gets modified which
affects the quantum phenomena. Recently, Bekenstein \cite{Bekenstein:2012yy,Bekenstein:2013ih}
proposed that quantum gravitational effects could be tested experimentally, he suggests ``a
tabletop experiment which, given state of the art ultrahigh vacuum and cryogenic technology,
could already be sensitive enough to detect Planck scale signals'' \cite{Bekenstein:2012yy}.
In a series of papers, the effects of GUP on atomic systems, condensed matter systems, preheating phase of the universe,
inflationary era of the universe and black holes have been  investigated
\cite{dvprl,dvcjp, Ali:2011fa,Chemissany:2011nq, Tawfik:2012he, Ali:2012mt, faragali, neutrinoGUP, Majumder:2012qy, Majumder:2012ph, Nozari:2012nf, Nozari:2012gd, majahep, majgrg, majahep2}.\par

In this paragraph we shed some light on the phase space formulation of quantum mechanics as in the later sections we are going to use GUP modified classical Poisson brackets. The known correspondence between commutator and Poisson bracket was first proposed by Dirac where he proposed that the quantum counterparts $\hat{A}$, $\hat{B}$ of classical observables $A$, $B$ satisfy \cite{1}
\bea
[\hat{A}, \hat{B}]= i \hbar \{A,B\}
\eea
and since that time, this is known as a postulate of Quantum Mechanics.
In 1946, Hip Groenewold demonstrated that a general systematic correspondence between quantum commutators and Poisson brackets could not hold consistently \cite{2}.  However, he did appreciate that such a systematic correspondence does, in fact, exist between the quantum commutator  and a deformation of the Poisson bracket,  which is called the Moyal bracket \cite{3}. The Moyal bracket is a way of describing the commutator of observables in the phase space formulation of quantum mechanics when these observables are described as functions on phase space. It relies on schemes for identifying functions on phase space with quantum observables, the most famous of these schemes being  Weyl quantization \cite{4}. It underlies Moyal's dynamical equation, an equivalent formulation of Heisenberg's quantum equation of motion, thereby providing the quantum generalization of Hamilton's equations. In a two-dimensional flat phase space, and for the Weyl-map correspondence, the Moyal  bracket reads,
\bea
\{\{f,g\}\} = \frac{1}{i\hbar} (f\star g- g\star f)= \{f,g\}+\mathcal{O}(\hbar^2).
\eea
Where $\star$  is the star-product operator in phase space (Moyal product), while $f$ and $g$ are differentiable phase-space functions, and $\{f,g\}$ is their Poisson bracket. This means Moyel bracket is equal to poisson bracket in its equivalence with the quantum commutator up to the second order of $\hbar$ (i.e $\hbar^2$). In this present paper we study the implications of GUP which has a linear term in $\alpha$ as well as a quadratic one. For the Friedmann model we only consider the correction which is proportional to $\alpha$ in the commutator of eqn.(\ref{comm01}) and $\hbar \alpha\sim \alpha_0 \ell_P\sim\hbar^{1/2}$ where $\ell_P=\sqrt{\hbar G/c^3}$, which means Moyel bracket is not necessary in our model and Poisson bracket is quite enough to study the commutator of GUP.
For studying Bianchi models we consider the most general form of GUP (including linear as well as quadratic term in $\alpha$) and also
GUP with only linear term in $\alpha$ in which we trust using Poisson bracket, and  we make a comparative analysis with earlier results in the literature \cite{montani}.

\par
Motivated by the GUP, many authors studied some classical problems with deformed Poisson brackets \cite{c36,c38}. From the phenomenological point of view, constraints were placed on deformation parameters in \cite{c36} by considering the effects of GUP motivated deformed Poisson algebra on the classical orbits of particles in the central force problem. In \cite{c38} it was conjectured that modified commutation relations as suggested by candidate theories of quantum gravity, persist in the classical limit also. There the perihelion precession rate for Keplerian orbits was calculated. A deformed Heisenberg algebra or in the classical case, a deformed Poisson algebra incorporate some additional problems like the violation of equivalence principle \cite{faragali}. But recently in \cite{c45} it was shown that the GUP is reconciled with the equivalence principle. The effects of the GUP on Galilean and Lorentz transformation is also studied recently in \cite{cmain}.\par
In this paper we are going to study the implication of the GUP as proposed in \cite{advplb} in early universe cosmology. The effect of $\alpha$ in (\ref{comm01}) is relevant in the very early universe or near Planck scale and it dies down at low energy scales and hence can be neglected at a later time in the evolution of the universe. However there are attempts to address the problem of Dark Energy in the realm of GUP. In \cite{barunMUP} the classical and quantum effects of this GUP are investigated on the phase space of a dilatonic cosmological model and it was found that it is possible to get a late time acceleration for this model. A similar study was done in \cite{darabi} and it was shown that GUP at first generates acceleration but prevents the eternal acceleration at late times and turns it into deceleration. The GUP can also incorporate corrections in the entropy area relation and thereby modify the energy densities in Holographic and Agegraphic Dark Energy Models. Due to this we can get distinct terms in the form of $f(R)$ which may have its importance in explaining the early inflationary scenario as well as the present accelerated expansion \cite{barunfr}.
\par
The scope of the present work is to investigate the effect of the GUP on the dynamics of the
Friedmann-Robertson-Walker universe (FRW) by performing this modification at the classical level by
studying the modifications induced on the symplectic geometry by the deformed algebra.
It is found that big bang singularity seems to be suppressed, by considering
GUP. It is found that  modified Friedmann equation predicts a cosmological bounce. Besides,
we extend our study to investigate the effect of GUP on Bianchi I and II universes.
In each case the deformed commutation relations due to GUP would modify the dynamical
equations and hence we will find departures from the usual scenario. We investigate the isotropy of Bianchi universes as well as
the singularity. Later we will see that the results remain qualitatively same compared to the GUP which has a quadratic term in $\alpha$ \cite{battisti}. However a discussion on the sign of $\alpha$ is made later which is important for the analysis.
An outline of this paper is as follows: in Sec. \ref{sec:FRW}, we investigate the GUP with FRW universe and derive the modified Freidmann equation. In Sec. \ref{Bianchi}, we tackle the impact of GUP on Bianchi universes with type I and II.
We compare our results with previous studies in \cite{montani}.

\section{FRW universe in the framework of GUP}
\label{sec:FRW}

Here we study the GUP deformed dynamics of an isotropic and homogeneous cosmological model. Let us start with a review of the standard case. The FRW metric can be described by the line element as
\bea
ds^2=- N^2 dt^2+a(t)^2 \le(\frac{dr^2}{1-k~r^2}+r^2 d\Omega^2\ri) \label{metric}
\eea
where $a(t)$ is the scale factor of the universe and $N$  is the lapse function. The constant $k$ is a measure of spatial flatness and it can be $0,\pm 1$. The dynamics of such models are summarized in the Hamiltonian constraint
\bea
\mathcal{H}=-\frac{2\pi G}{3} \frac{p_a^2}{a}-\frac{3}{8 \pi G} a k +a^3 \rho=0 \label{constraint}
\eea
where $G$ is the gravitational constant and  $\rho$ represents the matter density in the universe. The Freidmann equation can be derived using Poisson brackets
\bea
\{A,B\}=\le(\frac{\partial A}{\partial x_i}\frac{\partial B}{\partial p_j}- \frac{\partial A}{\partial p_i}\frac{\partial B}{\partial x_j}\ri) \{x_i,p_j\}\label{poisson}
\eea
where $x_i$ and $p_j$ are the canonical variables of the system. In the standard case, the canonical variables are given by $a$ and $p_a$  and these canonical variables satisfy the Poisson bracket as $\{a,p_a\}=1$. This is due to the fact that isotropy reduces the phase space of the model to two dimensional. The Friedmann equations can be extracted by the Hamilton's equations which can be derived from the extended Hamiltonian
\bea
\mathcal{H}_E=\frac{2\pi G}{3} \frac{N p_a^2}{a} + \frac{3}{8 \pi G} N a k - N a^3 \rho + \lambda \mathcal{P}
\eea
where $\lambda$ is the Lagrange multiplier, and $\mathcal{P}$ is the momentum conjugate to the lapse $N$ which vanishes. This would give the equation of motion for the lapse $\dot{N}= \{N,\mathcal{H}_E\}=\lambda$ and the scalar constraint of Eq. (\ref{constraint}) is obtained by the fact that the constraint $\mathcal{P} = 0$ will be satisfied at all times which is equivalent to demand that the secondary constraint $\dot{\mathcal{P}}= \{\mathcal{P},\mathcal{H}_E\}=\mathcal{H}=0 $ holds. The other equations of motion regarding $a$ and $p_a$ with respect to extended Hamiltonian are
\bea
\dot{a}&=&\{a,H_E\}=\frac{\partial H_E}{\partial p_a}= \frac{4 \pi G}{3} N \frac{p_a}{a} \label{dota} \\
\dot{p_a}&=& \{p_a,H_E\}=N\le(\frac{2 \pi G}{3} \frac{p_a^2}{a^2} -\frac{3}{8 \pi G} k+ 3 a^2 \rho +a^3 \frac{d\rho}{da}\ri)\label{dotp}
\eea
By using the equations (\ref{dota}) and (\ref{dotp}) and the scalar constraint (\ref{constraint}), we can obtain the equation of motion for $\dot{a}$ which represents the Freidmann equation
\bea
\le(\frac{\dot{a}}{a}\ri)^2= \frac{8 \pi G}{3} \rho - \frac{k}{a^2}
\eea
Usually $\dot{a}/a$ is called the Hubble rate. It is well known that this classical equation breaks down at the big-bang singularity and a natural crisis for a quantum description of the Universe in Planck scale occurs. \par
Now we use the GUP as proposed in \cite{advplb,Das:2010zf} for a heuristic analysis of the singularity at big-bang. Here we intend to study the classical equations only. As discussed earlier in details, here we will consider only the GUP motivated deformed Poisson algebra to study the classical equations that govern our universe. With implementing the  GUP up to the first order of  $\alpha$, we get the Poisson bracket between $a$ and $p_a$ as
\bea
\{a,p_a\}= 1-2 \a p_a ~~.
\eea
Here we have replaced the quantum mechanical commutator by the Poisson bracket. So, using the Poisson algebra of Eq. (\ref{poisson}), we get the modified equations of motions for FRW universe as
\bea
\dot{a}= \{a,\mathcal{H}_E\}= \frac{\partial H_E}{\partial p_a} (1-2 \a p_a)\\
\dot{p_a}=\{p_a,\mathcal{H}_E\}= - \frac{\partial H_E}{\partial a} (1-2 \alpha p_a)
\eea
Using the expression for the extended Hamiltonian we follow the same procedure as in the standard case and get the modified equations of motions as
\bea
\dot{a}&=& \frac{4 \pi G}{3} N \frac{p_a}{a} (1-2 \a p_a) \label{Mdota}\\
\dot{p_a}&=& \{p_a,H_E\}=N\le(\frac{2 \pi G}{3} \frac{p_a^2}{a^2} -\frac{3}{8 \pi G} k+ 3 a^2 \rho +a^3 \frac{d\rho}{da}\ri) (1-2\a p_a)\label{Mdotp}
\eea
Using equations (\ref{Mdota}) and (\ref{Mdotp}) with the scalar constraint (\ref{constraint}), we obtain the equation of motion for
the Hubble rate $\dot{a}/a$ which represents the modified/deformed Friedmann equation due to GUP as
\bea
\le(\frac{\dot{a}}{a}\ri)^2= \le(\frac{8 \pi G}{3} \rho - \frac{k}{a^2}\ri) \le[1- 2 \alpha a^2 \sqrt{\frac{3}{2 \pi G}}\le(\rho-\frac{3}{8 \pi G} \frac{k}{a^2}\ri)^{1/2}\ri]
\eea
By considering the spatially flat case in which k=0, we find that the modified Friedmann equation will be
\bea
\label{FR3}
\le(\frac{\dot{a}}{a}\ri)^2= \frac{8 \pi G}{3} \rho \le[1- 2 \alpha a^2 \sqrt{\frac{3}{2 \pi G}}\rho^{1/2}\ri]
\eea

This equation appears to represent a big bounce picture for the FRW universe with some critical density. This equation appears to introduce an interesting result, since it links the matter density $\rho$ with the expansion rate $a$. It implies that the value of matter density $\rho^{\frac{1}{2}}$ is always bounded with $1/(2 \alpha a^2 \sqrt{\frac{3}{2 \pi G}})$ so the correction due to GUP is always kept $\leq 1$, or otherwise we will get an imaginary density or expansion rate. This behavior is quite similar to the effect of GUP on black hole thermodynamics, in which the final stages of a black hole reach to a remnant due to GUP effect, and this remnant does not radiate any more, otherwise, we will get imaginary Hawking temperature \cite{Adler,Cavaglia:2003qk}. This lead us to define a critical density in the considered model as follows:
\bea
\rho_c^{1/2}= \frac{1}{2 \alpha a^2 \sqrt{\frac{3}{2 \pi G}}}\label{critden}
\eea
The inbuilt minimal length in the theory restricts the limit $a \rightarrow 0$ which keeps the \emph{dynamical}\footnote{By dynamical we mean the dependence of $\rho_c$ on the scale factor $a$.} value of $\rho_c$ finite.
Our modified Friedmann equation (\ref{FR3}) has a very close resemblance with results found in \cite{battisti}. In \cite{battisti} the modified Friedmann equations were studied with the form of the generalized uncertainty principle deduced from the Snyder non-commutative space. The resulting Poisson bracket used in the paper was $\{q,p\}=\sqrt{1-\alpha^{\prime}p^2}$ which is consistent with the one described in \cite{kmm}. There it was shown that the resulting cosmological model predicts a cosmological bounce at some critical energy density. Although our result is qualitatively similar to \cite{battisti} but the critical energy density in our case depends on the scale factor as seen in eqn.(\ref{critden}). Also due to the choice of the GUP which contains a linear term in the Planck length and momentum \cite{advplb} we get the modified Friedmann equation where the minimal length manifests itself in the $\rho^{3/2}$ term. But in \cite{battisti} the same dependence comes from the $\rho^2$ term because the deformed Poisson bracket had the minimal length with the quadratic term in momentum.\par
It is very important to discuss the sign and value of $\alpha$ in the GUP of eqn.(\ref{comm01}). The GUP is motivated by doubly special relativity (DSR) \cite{cg} where a non-linear realization of Lorentz transformations in energy-momentum space is parametrized by an invariant length scale $\alpha(=\frac{\alpha_0 l_{pl}}{\hbar})$. If {\it a priori} we do not consider $\alpha_0$ to be a order one parameter, at least for phenomenological purposes, then its existence may signal a new length scale which cannot exceed the electro-weak length scale $(\sim 10^{17}l_{pl})$ as otherwise its effects would have been observed \cite{Ali:2011fa}. This implies $\alpha_0 \leq 10^{17}$. $\alpha_0>0$ in the present analysis depicts a close resemblance of the modified Friedmann equation with the effective equation of LQC \cite{param}. For $\alpha_0<0$, the string inspired Randall-Sundrum braneworld scenario leads to a similar modification of the Friedmann equation \cite{maartens}. In this case we cannot get a bouncing scenario as $\dot{a}$ does not vanish at $\rho=\rho_c$. However if the extra dimension of the bulk spacetime is considered time-like then we get a similar prediction \cite{sahni}.\par
Near the Planck scale or at energy scales much greater than the electro-weak scale the term with $\alpha$ would be highly relevant and the vanishing of the Hubble rate may result in a cosmological bounce. At very low energy scales the term with $\alpha$ will be irrelevant and hence can be neglected giving back our standard Friedmann equation which is well suited to describe the later stages of the universe. A similar result was also found earlier in \cite{asphy} but interpreted in terms of the area of the apparent horizon. If we consider only the leading order correction due to GUP then it can be shown that $\dot{a}$ may vanish at some critical area of the apparent horizon.


\section{Bianchi Universes and GUP}
\label{Bianchi}

We now study the modified equations of motion for the Bianchi Universe by implementing the minimal length in Quantum gravity which assume a natural cut-off on the anisotropies. The Bianchi Universes are spatially homogeneous cosmological spacetimes and the symmetry group acts on each spatial manifold \cite{ryan}. In the Misner formalism \cite{Misner} the Hamiltonian constraint governing the dynamics can be written as
\begin{equation}
\label{H1}
H = -p_{\gamma}^2 + p_+^2 + p_-^2 + e^{4\gamma} V(x_{\pm})= 0 ~~,
\end{equation}
where the lapse function $N=-e^{3\gamma}/2p_{\gamma}$ is fixed by the time gauge $\dot{\gamma}=1$. $\gamma$ describes the isotropic expansion and the anisotropies are described by $x_{\pm}$. The classical singularity occurs in the limit $\gamma \rightarrow -\infty$ and the the potential term $V(x_{\pm})$ makes the classification between the models which is associated with the scalar curvature. Now it is necessary to make a choice of the time parameter for analyzing the dynamics of the system. Since the volume of the universe is proportional to $e^{3\gamma}$ so here we consider $\gamma$ to be the time parameter and obtain an effective Hamiltonian by solving the constraint equation (\ref{H1}) with respect to $p_{\gamma}$. The effective Hamiltonian can be written as
\bea
\label{H2}
\mathcal{H} = -p_{\gamma} = \le[p_+^2 + p_-^2 + e^{4\gamma} V(x_{\pm})\ri]^{\frac{1}{2}}
\eea
Let us now consider the modifications on the phase space as introduced by the GUP. Here we will use the GUP of \cite{advplb} for our purpose and according to it the Poisson brackets are
\bea
\{x_i,p_i\}&=& \delta_{ij}\hspace{-0.5ex} -
\hspace{-0.5ex} \alpha \hspace{-0.5ex}  \le( p \delta_{ij} +
\frac{p_i p_j}{p} \ri)
+ \alpha^2 \hspace{-0.5ex}
\le( p^2 \delta_{ij}  + 3 p_{i} p_{j} \ri) \hspace{-0.5ex} , \label{comm1}\\
\{p_i,p_j\}&=&\{x_i,x_j\}=0 \label{comm2}.
\eea
For any phase space function we have
\bea
\{A,B\}=\le(\frac{\partial A}{\partial x_i}\frac{\partial B}{\partial p_j} - \frac{\partial A}{\partial p_i}\frac{\partial B}{\partial x_j}\ri) \{x_i,p_j\}\label{poisson1}
\eea
Here we have considered the requirements that the GUP deformed Poisson brackets must also be bilinear, anti-symmetric, following the Leibniz rules and the Jacobi identity. The deformed classical dynamics of the Bianchi models can be obtained from the phase algebra of Eqs. (\ref{comm1}, \ref{comm2}, \ref{poisson1}). We
derive below the time dependence of anisotropies and their conjugate momenta using the effective Hamiltonian of Eq. (\ref{H2}) as
\bea
\{x_i,\mathcal{H}\} &=& \dot{x}_i = \frac{p_k}{H}\le(\delta_{ik}-\alpha(p \delta_{ij}+\frac{p_i~p_j}{p})+\alpha^2(p^2 \delta_{ik}+3 p_i p_j)\ri) ~~, \la{dotx}\\
\{p_i,\mathcal{H}\} &=& \dot{p}_i = -\frac{1}{2 \mathcal{H}} e^{4 \gamma} \le(\delta_{ik}-\alpha(p \delta_{ij}+\frac{p_i~p_j}{p})+\alpha^2(p^2 \delta_{ik}+3 p_i p_j)\ri)
\frac{\partial V}{\partial x_k}  ~~, \la{dotp1}
\eea
the differentiation in the last two equations is in terms of the time variable $\gamma$ and $p^2 $ is defined as $p^2= p_+^2 + p_-^2$.
The equations (\ref{dotp1}, \ref{dotx}) represent the modified  equations of motion for the homogeneous Universes due to GUP. In the next subsections, we are going to study in details the Binachi I and II universes based on this formalism.

\subsection{Bianchi I in the framework of GUP}

In this section we study the Bianchi models in the framework of the GUP. It is important to mention that we consider the most general form of GUP (with linear and also quadratic term in the Planck length and momentum) for our study. Later we will see that the effects of the two corrections are different. In each step we express our result for two different form of the GUP, one with only linear correction and one with both linear and quadratic corrections and finally compare our result with those of \cite{montani} which considers only the quadratic form in $\alpha$.\par
\begin{figure}[htb!]
\centering
\includegraphics[scale=0.35,width=7.cm,angle=0]{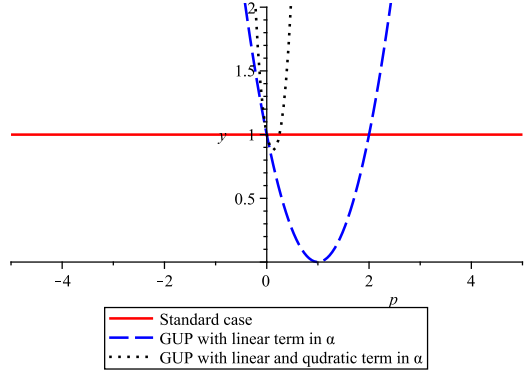}
\caption{$\dot{x}^2$ versus $p$ for standard and GUP modified case of linear term in $\alpha$ as well as linear and quadratic terms in $\alpha$}.
\label{solution}
\end{figure}

The main property of Bianchi I universe is to be homogeneous and it has flat spatial sections \cite{ryan, BianchiI}. Bianchi I corresponds to the case $V(x_{\pm})=0$ in the scheme of Eq. (\ref{H2}). This implies that the solution of the equations of motion (\ref{dotx}, \ref{dotp1})  of Bianchi I Universe is Kasner-like. Here we have
\bea
\dot{p}_i&=&0\\
\dot{x}^2&=& \frac{p^2}{\mathcal{H}^2} \le[1-4\alpha p+12 \alpha^2 p^2-16 \alpha^3 p^3+16 \alpha^4 p^4\ri]\nn\\
&=& \le[1-4\alpha p+12 \alpha^2 p^2-16 \alpha^3 p^3+16 \alpha^4 p^4\ri] \nn \\ &=& [1-2 \alpha p+ 4 \alpha^2 p^2]^2
\label{xdot}
\eea
For GUP with only a linear term in $\alpha$ the equation for $\dot{x}^2$ can be written as
\bea
\dot{x}^2 = [1-2\alpha p]^2
\eea
As $\alpha=0$, we get standard Kasner velocity which is $\dot{x}^2=1$. The effect of GUP and its cutoff enhance the values Kasner velocity. In the last step, we used  the fact for the Bianchi I universe, the ADM Hamiltonian is constant and given by $\mathcal{H}^2 = p^2 = \text{const}$. In Fig. (\ref{solution}) we plot $\dot{x}^2$ as a function of $p$ for GUP modification and standard cases.

We investigate here how the Kasner behavior is affected by the GUP deformed framework. The spatial metric of the Kasner solution is written as
\bea
dl^2 = t^{2s_1} dx_1^2 + t^{2s_2} dx_2^2 + t^{2s_3} dx_3^2 ~~,
\eea
where $s_1, s_2, s_3$ are the Kasner indices which follow the following equations:
\bea
s_1^2 + s_2^2 + s_3^2 = 1\\
s_1 + s_2 + s_3 = 1 ~~.
\eea
The first Kasner relation is directly related to the anisotropy velocity $\dot{x}$ by the equations \cite{BianchiI}
\bea
\dot{x}_+ &=& \frac{1}{2} (1-3 s_3) \\
\dot{x}_- &=& \frac{\sqrt{3}}{2} (s_1 - s_2) ~~.
\eea
and the second arises from the arbitrariness in choosing the tetrads and is still valid in the GUP deformed formulation. Then, the first Kasner relation is then deformed as
\begin{align}
\label{seqn}
s_1^2 + s_2^2 + s_3^2 &= 1 - 4\alpha p + 12 \alpha^2 p^2 - 16 \alpha^3 p^3 + 16 \alpha^4 p^4 \nonumber \\
 &= 1 - 4\sqrt{\mu} + 12 \mu - 16 \mu^{3/2} + 16 \mu^2 ~.
\end{align}
For GUP with only a linear term in $\alpha$ this equation can be written as
\bea
s_1^2 + s_2^2 + s_3^2 = 1 - 4\sqrt{\mu} + 4 \mu \label{slinear}
\eea
Here we have defined $\mu=\alpha^2 p^2$ as a measure of deformation in terms of the anisotropy momentum. $\mu=0$ gives the standard result. Now it is very important to note that in the GUP framework that we are using the terms on the right hand side of Eq. (\ref{seqn}) have alternating signs. This means that the Universe can isotropize and can reach a situation when the Kasner indices become equal. But in the framework of GUP as introduced by Kempf et. al. \cite{kmm} the Universe cannot isotropize and the Kasner dynamics is highly modified which is just the opposite as in our case \cite{montani}. So in our case we can accommodate the contraction along two directions while approaching the classical singularity and is similar to what happens in the standard case.
On the other hand, for the GUP with a linear term in $\alpha$, the terms on the right-hand side of (\ref{slinear}) have mixed signs, which means that the Universe can isotropize, i.e. it can reach the stage where the Kasner indices are equal. The form of GUP that we have considered comes with a characteristic length scale $\alpha$ and its presence is linear as well as quadratic in the commutation relation. Generally we consider it to be positive but if $\alpha <0$ then our result will have no qualitative difference with the result found in \cite{montani}. To capture the GUP effects in detail we need to investigate the GUP deformed Bianchi II Universe.

\subsection{Bianchi II in the framework of GUP}

\begin{figure}[htb]
\centering
\begin{tabular}{c}
\hspace{2cm} \includegraphics[scale=0.9,width=9.cm,angle=0]{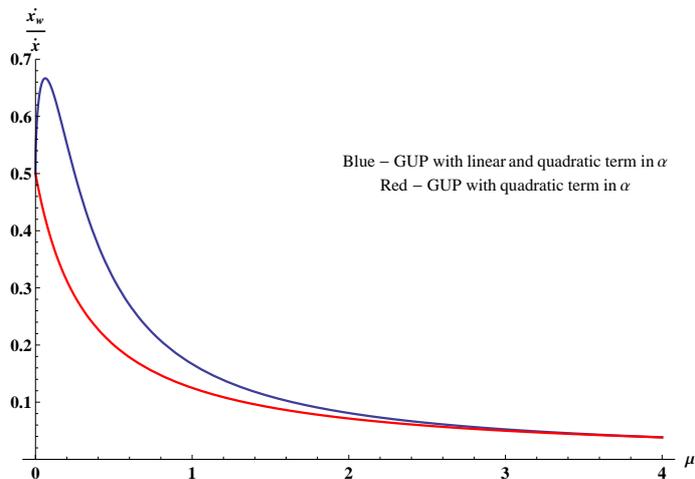}
\end{tabular}
\caption{\footnotesize Plot of $\dot{x}_w/\dot{x}$ as a function of $\mu$. Here we have compared our result (blue) with the result of the GUP with a quadratic term in $\alpha$ \cite{montani} (red).}
\label{fig2}
\end{figure}

In this section we will study Bianchi II dynamics in the same framework of the GUP. Bianchi II connects the homogeneous flat Universe of that of Bianchi I with Bianchi IX. Bianchi II corresponds to Bianchi IX when we consider only one potential wall \cite{BianchiI,ryan}. As we are considering the Hamiltonian analysis, for Bianchi II the potential is as follow; $V(x_{\pm})=e^{-8x_+}$. We can write the Hamiltonian as
\begin{equation}
\mathcal{H} = [p_+^2 + p_-^2 + e^{4(\gamma - 2 x_+)}]^{\frac{1}{2}} ~~.
\end{equation}
The main difference of the GUP framework with respect to the standard one is that $\mathcal{H}$ is not anymore a constant of motion in the vicinity of the classical singularity $\gamma \rightarrow -\infty$. We need to investigate the bounce of the Universe (or analogous to particle) against the potential wall in the GUP framework. For that we need to have the expression of wall velocity. In this case the wall velocity is written as follows \cite{BianchiI,montani}
\begin{equation}
\dot{x}_+ \approx \dot{x}_w = \frac{1}{2} - \frac{1}{8} ~\frac{\partial}{\partial \gamma} ~\ln \mathcal{H}^2 ~~.
\end{equation}
Here in the GUP framework $\mathcal{H}$ is not a constant and we can write
\begin{equation}
\frac{\partial}{\partial \gamma} ~\ln \mathcal{H}^2 = 4 \left[1 - \frac{p^2(\dot{x})}{\mathcal{H}^2}\right] ~~.
\end{equation}
So the wall velocity now becomes
\bea
\label{wallvelo1}
\dot{x}_w &=& \frac{p^2(\dot{x})}{2\mathcal{H}^2} = \frac{\dot{x}^2}{2}[1 - 4 \alpha p + 12 \alpha^2 p^2 - 16\alpha^3 p^3 + 16 \alpha^4 p^4]^{-1} \nn \\&=&
\frac{\dot{x}^2}{2} \le[1-2 \a p +4 \a^2 p^2\ri]^{-2}
\eea
Here we have used Eq. (\ref{xdot}) for $\dot{x}$ and considered that $\mathcal{H}^2=p^2$ near the classical cosmological singularity ($\gamma \rightarrow -\infty$). Eq. (\ref{wallvelo1}) can further be written as
\begin{equation}
\frac{\dot{x}_w}{\dot{x}} = \frac{1}{2}(1 - 2 \sqrt{\mu} + 4 \mu)^{-1} ~~,
\end{equation}
where we have defined $\mu=\alpha^2 p^2$. For GUP with only a linear term in $\alpha$ this equation can be written as
\bea
\frac{\dot{x}_w}{\dot{x}} = \frac{1}{2}(1 - 2 \sqrt{\mu})^{-1} \label{linearwall}~~.
\eea
In the undeformed state $\mu=0$, we recover the standard picture where $\dot{x}=1$ and $\dot{x}_w=1/2$.
In Fig. (\ref{fig2}) we plot the ratio of wall velocity and particle velocity as a function of $\mu$. We can see that in the GUP framework that we are studying, the ratio $\dot{x}_w/\dot{x}$ is higher and have a maximum for low values of $\mu$ or in some sense in the quasi standard regime. The result converges with the result of the GUP framework of \cite{kmm} for high values of $\mu$ or in the highly deformed regime. In our case we can clearly see that the bounce is accelerated for low values of $\mu$ as there is maximum of $\dot{x}_w/\dot{x}$ at around $0.67$.

In the GUP framework the particle (Universe) as well as the wall velocity depends on
the anisotropy momentum and the deformation parameter $\alpha$. Here
$\dot{x}_w/\dot{x} \neq 1/2$ as what is expected in the standard case.
In the deformed case in the asymptotic limit of very large $\mu$, $\dot{x}_w/\dot{x}$
vanishes and the wall appears stationary with respect to the particle (Universe) velocity.
The maximum angle for the bounce to happen is evaluated as $\vert \theta_{max}\vert = \pi/3$ in
the standard case \cite{mon21}. In the asymptotically limit of the highly deformed
case ($\mu>>1$) the maximum angle is evaluated as
$\vert \theta_i \vert < \vert \theta_{max} \vert =\cos^{-1} (\dot{x}_w/\dot{x}) = \pi/2$.
Based on this, when the momentum of the particle (Universe) evolve into a maximum measurable energy, the bounce is
enhanced against the wall which means no maximum limit angle is manifested anymore. Also we have a maximum of $\dot{x}_w/\dot{x}$ for low values of $\mu$ at around $0.67$ where the phenomenon of bounce is accelerated. The maximum angle for this case is $\vert \theta_{max}\vert \approx \frac{4\pi}{15}$. \par
\begin{figure}[htb]
\centering
\begin{tabular}{c}
\hspace{2cm} \includegraphics[scale=0.9,width=9.cm,angle=0]{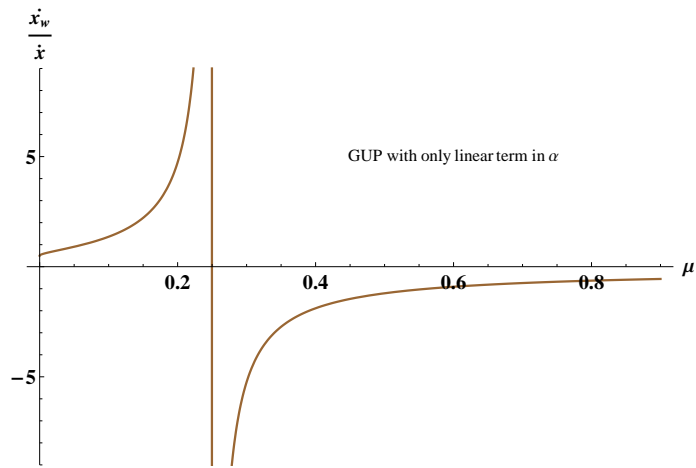}
\end{tabular}
\caption{\footnotesize Plot of $\dot{x}_w/\dot{x}$ as a function of $\mu$ for the GUP with only a linear term in $\alpha$.}
\label{fig3}
\end{figure}

It is interesting to note that for GUP with only a linear term in $\alpha$, $\dot{x}_w/\dot{x}$ has a discontinuity at some lower value of $\mu$ but vanishes asymptotically at larger $\mu$ with negative values (\ref{linearwall}). In Fig. (\ref{fig3}) we plot the ratio of wall velocity and particle velocity as a function of $\mu$. Here the wall appears moving in the opposite direction with respect to the particle (Universe) velocity for large values of $\mu$. At some value of $\mu$ there is a discontinuity and the particle (Universe) velocity seems stationary. Earlier we have mentioned that the form of GUP that we have considered comes with a characteristic length scale $\alpha$ and generally we consider it to be positive. But if $\alpha <0$ then our result will have no qualitative difference with the result found in \cite{montani}.

\section{Conclusions}
In this work we have investigated the consequences of the GUP, which
predicts maximum observable momenta besides a minimal measurable length \cite{advplb,Das:2010zf} and is
consistent with DSR\cite{cg}, string theory and black hole physics \cite{guppapers, BHGUP},
on FRW universe and Bianchi I and II universes. By investigating the effect of GUP with FRW universe,
we found that the big bang singularity seems to be suppressed by a dynamical critical density (dynamical in the sense that it is scale factor dependent). Moreover, we extend our study for Bianchi I and II universes with different forms of GUP available in the literature. By investigating different forms of GUP with Bianchi I, we found that the Universe may possibly be isotropic and may evolve into a situation at which the Kasner indices become equal with implementing GUP. With Bianchi II, we found that  in the asymptotic limit of the highly deformed
case ($\mu>>1$) the maximum angle is given by
$\vert \theta_i \vert < \vert \theta_{max} \vert =\cos^{-1} (\dot{x}_w/\dot{x}) = \pi/2$.
when the momentum of the particle (Universe) evolves into a maximum measurable energy, the bounce is
enhanced against the wall which means no maximum limit angle is manifested anymore. For low $\mu$,  we have a maximum of $\dot{x}_w/\dot{x}$ for low values of $\mu$ at around $0.67$ where the phenomenon of bounce is accelerated. The maximum angle for this case is $\vert \theta_{max}\vert \approx \frac{4\pi}{15}$. We conclude that the proposed GUP in this work might possibly resolve singularity problems with the considered universes and may imply a bounce picture for the universe.\par
We should also point out that an expanding universe becomes isotropic due to the matter contributions. But here we have not considered any particular matter which is a shortcoming of our present model. So it is indeed necessary to study the effective equations of our model with different matter contributions which we plan to consider in near future.

\section*{Acknowledgments}
The authors would like to thank an anonymous referee for enlightening comments and helpful suggestions.
The research of AFA is supported by Benha University (www.bu.edu.eg) and CTP in Zewail City. AFA would like
to thank ICTP, Trieste for the kind hospitality where the present work was started.


\end{document}